\documentclass[twocolumn,amsmath,amssymb,superscriptaddress,aps,pra]{revtex4-1}%superscriptaddress,
\usepackage{amsmath,amsthm,amssymb}
\usepackage{latexsym}
\usepackage{amscd,graphicx}
\usepackage[titletoc]{appendix}
\usepackage{epsfig}
\usepackage{epstopdf}
\usepackage[normalem]{ulem}
\usepackage[dvipsnames]{xcolor}
\usepackage[colorlinks=true,linkcolor=blue,citecolor=red]{hyperref}
\usepackage{csquotes}

\begin{document}
	
\title{Nonreciprocal Microwave-Optical Entanglement in Kerr-Modified Cavity Optomagnomechanics}

\author{Ming-Yue Liu}%(刘明月)
\affiliation{Department of Physics, Wenzhou University, Zhejiang 325035, China}

\author{Yuan Gong}%(龚媛)
\affiliation{Department of Physics, Wenzhou University, Zhejiang 325035, China}

\author{Jiaojiao Chen}%(陈姣姣)
%\altaffiliation{jjchenphys@hotmail.com}
\affiliation{Department of Physics, Wenzhou University, Zhejiang 325035, China}
\affiliation{School of Physics and Optoelectronics Engineering, Anhui University, Hefei 230601, China}

\author{Yan-Wei Wang}%(王艳伟)
\altaffiliation{wangyw@wzu.edu.cn}
\affiliation{Department of Physics, Wenzhou University, Zhejiang 325035, China}

\author{Wei Xiong}%(熊伟)
\altaffiliation{xiongweiphys@wzu.edu.cn}
\affiliation{Department of Physics, Wenzhou University, Zhejiang 325035, China}
\affiliation{International Quantum Academy, Shenzhen, 518048, China}

\date{\today }

\begin{abstract}
Microwave-optical entanglement is essential for efficient quantum communication, secure information transfer, and integrating microwave and optical quantum systems to advance hybrid quantum technologies. In this work, we demonstrate how the magnon Kerr effect can be harnessed to generate and control nonreciprocal entanglement in cavity optomagnomechanics (COMM). This effect induces magnon frequency shifts and introduces pair-magnon interactions, both of which are tunable through the magnetic field direction, enabling nonreciprocal behavior. By adjusting system parameters such as magnon frequency detuning, we show that magnon-phonon, microwave-optical photon-photon, and optical photon-magnon entanglement can be nonreciprocally enhanced and rendered more robust against thermal noise. Additionally, the nonreciprocity of entanglement can be selectively controlled, and ideal nonreciprocal entanglement is achievable. This work paves the way for designing nonreciprocal quantum devices across the microwave and optical regimes, leveraging the unique properties of the magnon Kerr effect in COMM.\\


\end{abstract}

\maketitle

\section{introduction}

Magnons~\cite{rameshti2022cavity,yuan2022quantum,prabhakar2009spin,van1958spin}, the collective excitations of spins in magnetically ordered materials, are attracting increasing interest in quantum information science. Among these materials, yttrium-iron-garnet (YIG, $\mathrm{Y}_3 \mathrm{Fe}5 \mathrm{O}{12}$)\cite{schmidt2020ultra,mallmann2013yttrium,geller1957crystal} is particularly notable for its high spin density and low loss\cite{li2020hybrid}. In YIG, magnons can strongly couple with microwave photons in cavities~\cite{huebl2013high,tabuchi2014hybridizing,li2020hybrid,zhang2014strongly}, enabling the exploration of phenomena such as dark modes~\cite{zhang2015magnon}, non-Hermitian exceptional points~\cite{zhang2017observation,zhang2019experimental,zhao2020observation,sadovnikov2022exceptional,liu2019observation,cao2019exceptional}, one-way quantum steering~\cite{yang2021controlling}, dissipative coupling~\cite{wang2019nonreciprocity,harder2021coherent}, spin-photon interaction~\cite{hei2021enhancing}, and near-perfect absorption~\cite{rao2021interferometric}. Recently, cavity magnonics has entered the true quantum regime with the realization of single-magnon manipulation through coupling YIG and a superconducting qubit to a three-dimensional microwave cavity~\cite{xu2023quantum,xu2024macroscopic}.

Magnons can also couple with phonons via the magnetostrictive effect, forming the basis of cavity magnomechanics~\cite{zuo2024cavity}. This coupling mirrors the radiation pressure mechanism in optomechanics~\cite{aspelmeyer2014cavity}, allowing the direct transfer of optomechanical phenomena~\cite{aspelmeyer2014cavity,vitali2007optomechanical,xiong2022higher,xiong2021higher,xiong2016cross,kani2022intensive} into the magnomechanical domain. Magnomechanics further enables the study of tripartite entanglement~\cite{li2018magnon}, squeezing~\cite{li2019squeezed}, and non-Hermitian physics~\cite{lu2021exceptional,he2024mechanical}. Extending this framework, the mechanical motion in magnomechanics can couple to an optical cavity via radiation pressure, giving rise to cavity optomagnomechanics (COMM)~\cite{fan2023microwave,fan2022optical}. This hybrid platform provides a versatile foundation for exploring diverse quantum phenomena~\cite{zuo2024cavity}.

In addition, the Kerr effect of magnons (i.e., Kerr magnons), stemming from the magnetocrystalline anisotropy in YIG~\cite{zhang2019theory}, was experimentally observed~\cite{wang2016magnon,wang2018bistability}, paving the way for investigating Kerr-modified quantum phenomena~\cite{zheng2023tutorial,shen2022mechanical}. This nonlinearity facilitates the study of multistability~\cite{shen2022mechanical,bi2021tristability}, spin-spin interactions~\cite{xiong2022strong,xiong2023optomechanical,peng2023strong}, and quantum phase transitions~\cite{zhang2021parity,liu2023switchable}. These findings emphasize the potential of magnons in hybrid quantum systems, where multiple quantum subsystems can be coupled for advanced quantum control and information processing. Additionally, Kerr magnons may be used to explore continuous variable entanglement~\cite{chen2024nonreciprocal,chen2023nonreciprocal}, although such entanglement has been largely confined to the microwave frequency domain. For broader quantum information applications, optical photons (or flying qubits) are preferred. Thus, achieving microwave-optical entanglement becomes critical. {Various proposals have been put forward in the context of hybrid electro-optomechanics~\cite{zhong2020entanglement, barzanjeh2012reversible, wu2021deterministic, zhong2020proposal, meesala2024quantum, wei2022tunable, sahu2023entangling}, where optical and microwave photons are coupled to a common mechanical motion via radiation pressure and capacitance change. Recently, microwave-optical entanglement has been extended to cavity magnomechanics through magnetostrictive forces~\cite{zheng2024nonreciprocal, zuo2023cavity, li2025microwave}.} The directionality dependence of the magnon Kerr effect in experiments~\cite{wang2016magnon,wang2018bistability} suggests that different magnetic field orientations can lead to opposite Kerr effects, which may introduce nonreciprocal behaviors in magnon-related hybrid systems~\cite{chen2024nonreciprocal,chen2023nonreciprocal,fan2024nonreciprocal}. Previous studies have demonstrated that nonreciprocity can enhance and protect continuous variable entanglement~\cite{jiao2020nonreciprocal}. This raises the possibility that nonreciprocal magnon Kerr effects could enhance microwave-optical entanglement, although this has not yet been explored.

In this work, we propose a scheme for utilizing the magnon Kerr effect to generate and control nonreciprocal microwave-optical entanglement in COMM. The proposed system consists of a magnonic nanoresonator that is simultaneously coupled to both {an optical cavity and a magnon mode via the radiation pressure and magnetostrictive force, respectively}. The magnonic resonator is embedded within the microwave cavity, facilitating photon-magnon interactions. The optical cavity is constructed using a fixed mirror and a highly reflective, small, and lightweight mirror attached to the magnonic nanoresonator. Under strong driving fields, the magnon Kerr effect induces a frequency shift in the magnons, which can be either positive or negative, similar to the Sagnac effect~\cite{maayani2018flying,malykin2000sagnac}. This effect also gives rise to a pair-magnon interaction that regulates the values of all entanglement. Both the frequency shift and the coefficients of the pair-magnon interaction are sensitive to the direction of the external magnetic field, leading to nonreciprocal behavior in the entanglement. By tuning system parameters, such as magnon frequency detuning, we find that nonreciprocal enhancement of magnon-phonon, microwave-optical, and optical photon-magnon entanglements is possible when the magnetic field is changed. Furthermore, we demonstrate that ideal nonreciprocity of microwave-optical and optical photon-magnon entanglements is achievable through the defined bidirectional contrast ratio. This type of entanglement, transitioning from the microwave to the optical band, has broad applications in quantum information processing, particularly in quantum networks~\cite{schoelkopf2008wiring,pirandola2016physics,krastanov2021optically,agusti2022long} and hybrid quantum systems~\cite{xiang2013hybrid,clerk2020hybrid}. Our work offers a promising avenue to  employ the magnon Kerr effect for designing nonreciprocal quantum devices that operate across both microwave and optical bands in COMM. {Note that our work differs from previous studies~\cite{zheng2024nonreciprocal, zuo2023cavity, li2025microwave}. Compared to works~\cite{zuo2023cavity, li2025microwave}, we employ an LC resonator rather than a 3D cavity. Additionally, magnons with the Kerr effect are introduced to investigate nonreciprocity, which distinguishes our approach from the study in Ref.~\cite{zheng2024nonreciprocal}, where the nonreciprocity arises from the Sagnac effect of photons.}

This paper is organized as follows. In Sec.~\ref{s2}, the model and the system Hamiltonian are described.  Then the dynamics of the system  and the quantity of the bipartite entanglement are given in Sec.~\ref{s3}. In Sec.~\ref{sec4}, the generation of nonreciprocal microwave-optical entanglement is studied. Finally, a conclusion is given in Sec.~\ref{s5}.

\section{MODEL AND HAMILTONIAN}\label{s2}
\begin{figure}
	\includegraphics[scale=0.6]{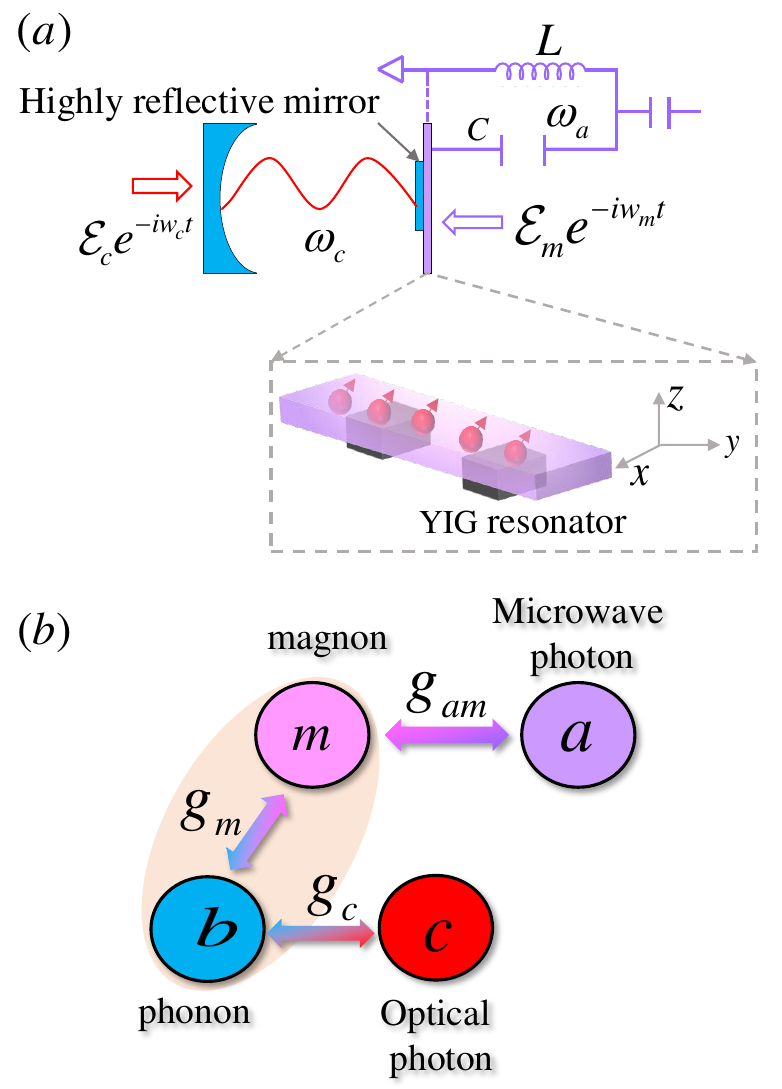}  
	\caption{(a) Schematic of the Kerr-modified COMM system consisting of an optical cavity, a YIG nanoresonator (see the bridge structure in the dashed box) and a microwave cavity (LC resonator), where the optical cavit is formed by two mirrors, one is fixed, the other is a highly reflective mirror attached to the YIG resonator. The YIG nanoresonator supports the Kerr magnons and phonon mode due to the magnetostrictive effect. The magnon Kerr effect stems from the magnetocrystallographic
anisotropy, tuned by the magnetic field. $\mathcal{E}_{c(a)}$ is the driving field with frequency $w_{c(a)}$. (b) The coupling configuration. The mechanical mode is simultaneously coupled to both the optical photons and microwave magnons via the radiation pressure and magnetostrictive force, respectively. The coupling strengths are $g_c$ and $g_m$. The magnons also interact the microwave phonons with the coupling strength $g_{\rm am}$.}\label{fig1}
\end{figure}
We consider a Kerr-modified COMM system comprising a microwave LC resonator (resonance frequency $\omega_a$), a magnonic nanoresonator, and an optical cavity (resonance frequency $\omega_c$), as illustrated in Fig.~\ref{fig1}(a). The magnonic nanoresonator is modeled as a micro-scale bridge supporting both a magnon mode $m$ (resonance frequency $\omega_m$) with Kerr nonlinearity $K_0$ and a mechanical mode (resonance frequency $\omega_b$). Through the magnetostrictive effect~\cite{zhang2016cavity}, the magnon mode dispersively couples with the mechanical mode, forming a magnomechanical interaction. Additionally, the magnon mode is coupled to the microwave LC resonator, while the mechanical mode interacts dispersively with the optical cavity through radiation pressure~\cite{aspelmeyer2014cavity}. The coupling mechanisms are schematically represented in Fig.~\ref{fig1}(b). {Note that the LC resonator can be modeled by either a coplanar-waveguide resonator or a transmission-line resonator. Experimentally, magnons in YIG materials have been strongly coupled to photons in the planar cavity structure~\cite{huebl2013high}.} Under the application of two driving fields, one to the optical cavity and the other to the magnon mode, the total Hamiltonian of the hybrid system can be expressed as (assuming $\hbar=1$):
\begin{align}
	H=H_0+H_{\rm om}+H_{\rm mm}+H_{\rm km}+H_{\rm drive},\label{eq1}
\end{align}
where
\begin{align}
	H_0=&\omega_a a^{\dagger} a+\omega_c c^{\dagger} c+\omega_m m^{\dagger} m+\frac{\omega_b}{2}\left(q^2+p^2\right),\notag\\
	H_{\rm om}=&-g_c c^{\dagger} c q,~~H_{\rm mm}=g_m m^{\dagger} m q,\notag\\
	H_{\rm km}=&g_{\rm am} (a^{\dagger} m+m^{\dagger} a)+K_0 (m^{\dagger} m)^2,\notag\\
	H_{\rm drive}=&i \mathcal{E}_m m^{\dagger} e^{-i w_m t}+i \mathcal{E}_c c^{\dagger} e^{-i w_c t}+{\rm H.c.}.\notag
\end{align}
The first term, $H_0$, represents the free energy of the system. The second term, $H_{\rm om}$, denotes the optomechanical interaction between the mechanical mode and the optical cavity, characterized by the single-photon coupling strength $g_c$. The third term, $H_{\rm mm}$, describes the magnomechanical interaction between the mechanical mode and the magnon mode, with the single-magnon coupling strength $g_m$. The fourth term, $H_{\rm km}$, accounts for the coupling between the microwave LC resonator and the Kerr magnon mode, with a coupling strength $g_{\rm am}$. Experimentally, $g_{\rm am}$ can reach the strong coupling regime, while the Kerr nonlinearity $K_0$ is tunable from negative to positive by varying the direction of the magnetic field~\cite{wang2018bistability}. The final term, $H_{\rm drive}$, represents the Hamiltonian of the two driving fields, where $\mathcal{E}_c$ and $\mathcal{E}_m$ are the amplitudes and $w_c$ and $w_m$ are the frequencies. In Eq.~(\ref{eq1}), $q$ and $p$ denote the two quadratures of the mechanical resonator.

\section{DYNAMICS AND ENTANGLEMENT METRIC}\label{s3}

When dissipation is included, the dynamics of the system can be governed by the quantum Langevin equations. In the rotating frame with respect to two driving fields, we have
\begin{align}\label{eq2}
	\dot{q}=&\omega_b p,\notag \\
	\dot{p}=&-\omega_b q-\gamma_b p+g_c c^{\dagger} c-g_m m^{\dagger} m+\xi,\notag \\
	\dot{a}=&-\left(\kappa_a+i \Delta_a\right) a-i g_{\rm am} m+\sqrt{2 \kappa_a} a_{\rm i n},\notag\\
	 \dot{c}=&-(\kappa_c+i \Delta_c) c+i g_c c q+\mathcal{E}_c+\sqrt{2 \kappa_c} c_{\rm i n}, \\
	 \dot{m}=&-(\kappa_m+i \Delta_m) m-i g_m m q+\mathcal{E}_m+\sqrt{2 \kappa_m} m_{\rm i n} \notag\\
	 &-i g_{\rm am} a-2 i K_0 m^{\dagger} m m,\notag
\end{align}
Here, $\Delta_{a(m)} = \omega_{a(m)} - w_m$ represents the detuning of the LC resonator (magnon mode) from the driving field with frequency $w_m$, while $\Delta_c = \omega_c - w_c$ is the detuning of the optical cavity from the driving field with frequency $w_c$. The decay rates of the LC resonator, optical cavity, magnon mode, and mechanical resonator are denoted by $\kappa_a$, $\kappa_c$, $\kappa_m$, and $\gamma_b$, respectively. $\sigma_{\text{in}}$ ($\sigma = a, c, m$) are the input noise operators satisfying
\begin{align} \langle \sigma_{\text{in}}(t) \rangle =& 0,\notag\\
\langle \sigma_{\text{in}}^{\dagger}(t) \sigma_{\text{in}}(t') \rangle = &n_\sigma(\omega_\sigma) \delta(t - t'),\\
\langle \sigma_{\text{in}}(t) \sigma_{\text{in}}^{\dagger}(t') \rangle= &\left[n_\sigma(\omega_\sigma) + 1\right] \delta(t - t'), \notag 
\end{align}
where $n_\sigma(\omega_\sigma) = \left[\exp\left(\hbar \omega_\sigma / k_B T\right) - 1\right]^{-1}$ is the mean thermal excitation number at frequency $\omega_\sigma$, $k_B$ is the Boltzmann constant, and $T$ is the bath temperature.

The mechanical resonator is also subject to a Brownian stochastic force $\xi(t)$, which is inherently non-Markovian. However, for a high mechanical quality factor $Q_b = \omega_b / \gamma_b \gg 1$, the force can be approximated as Markovian~\cite{benguria1981quantum,giovannetti2001phase}. In this regime, $\xi(t)$ obeys
\begin{align}
{\langle \xi(t) \xi(t') + \xi(t') \xi(t) \rangle} \simeq 2\gamma_b \left[2 n_b(\omega_b) + 1\right] \delta(t - t'), 
\end{align}
where $n_b(\omega_b)$ is the mean thermal excitation number of the mechanical resonator, given by $n_b(\omega_b) = \left[\exp\left(\hbar \omega_b / k_B T\right) - 1\right]^{-1}$.

For long-time evolution, the system reaches its steady state, which implies that all time derivatives in Eq.~(\ref{eq2}) vanish.  By assuming that $\langle \cdot\rangle$ denotes the steady-state value of the operator, we then have $\langle a\rangle = -{g_{\rm am}\langle m\rangle}/{\Delta_a}$, $\langle c\rangle = -{i \mathcal{E}_c}/{\tilde{\Delta}_c}$, $\langle p\rangle=0$, $\langle q\rangle ={(g_c |\langle c\rangle|^2 - g_m |\langle m\rangle|^2)}/{\omega_b}$, and 
\begin{align} \label{q5}
	\langle m\rangle =& \frac{i \mathcal{E}_m \Delta_a}{g_{\rm am}^2 - \left(\tilde{\Delta}_m + \Delta_k\right)\Delta_a},
\end{align}
where $\tilde{\Delta}_m = \Delta_m + g_m\langle q\rangle$ and $\tilde{\Delta}_c = \Delta_c - g_c\langle q\rangle$ are effective detunings induced by the displacement of the mechanical resonator, $\Delta_k = 2 K_0 |\langle m\rangle|^2$ represents the frequency shift induced by the magnon Kerr effect. In particular, $\Delta_k$ can be positive or negative, depending on the sign of $K_0$, which can be easily tuned by changing the magnetic field. To obtain above results, we have assumed that the decay rate is much smaller than its corresponding effective frequency detuning.

With steady-state values in hand, we expand each operator around its steady-state value, i.e., 
\begin{align}
	\sigma=\langle \sigma\rangle+\delta\sigma,~~q=\langle q\rangle +\delta q,~~p=\langle p\rangle+\delta p,
\end{align}
where the symbol $\delta$ denotes the meaning of fluctuation. Substutiting these expressions back into Eq.~(\ref{eq2}) and neglecting higher-order fluctuation terms ensured by the strong driving fields (e.g., $\left|\langle m \rangle\right|, \left|\langle c \rangle\right| \gg 1$)~\cite{vitali2007optomechanical}, the linearized dynamics for the fluctuations can be reduced to
\begin{align}\label{eq4}
	\delta \dot{q} =& \omega_b \delta p, \notag \\
	\delta \dot{p} =& -\omega_b \delta q - \gamma_b \delta p - i G_c\left(\delta c^\dagger - \delta c\right) / \sqrt{2} \notag \\ 
	&- i G_m\left(\delta m^\dagger - \delta m\right) / \sqrt{2} + \xi, \notag \\
	\delta \dot{a} =& -\left(\kappa_a + i \Delta_a\right) \delta a - i g_{\rm am} \delta m + \sqrt{2 \kappa_a} a_{\rm in}, \\
	 \delta \dot{m} =& -\left[\kappa_m + i\left(\tilde{\Delta}_m + 2 \Delta_k\right)\right] \delta m - i g_{\rm am} \delta a \notag \\
	  &+ G_m \delta q / \sqrt{2} +i \Delta_k \delta m^\dagger + \sqrt{2 \kappa_m} m_{\rm in}, \notag \\
     \delta \dot{c} =& -\left(\kappa_c + i \tilde{\Delta}c\right) \delta c + G_c \delta q / \sqrt{2} + \sqrt{2 \kappa_c} c_{\rm in}, \notag
\end{align}
where $G_c = i \sqrt{2} g_c \langle c\rangle$ is the enhanced optomechanical coupling by multi-photon effect, and $ G_m = -i \sqrt{2} g_m \langle m\rangle$ is the enhanced magnomechanical coupling by multi-magnon effect.

By further introducing quadratures for the LC resonator, optical cavity and the magnon mode, i.e., $\delta X_\sigma=\left(\sigma+\sigma^{\dagger}\right) / \sqrt{2}$ and $\delta Y_\sigma=\left(\sigma-\sigma^{\dagger}\right) / i\sqrt{2}$, the dynamics in Eq.~(\ref{eq4}) can be equivalently written as the matrix form,
\begin{align}\label{eq5}
	\dot{u}(t)=A u(t)+n(t),
\end{align}
where $u(t)= [\delta X_a(t), \delta Y_a(t), \delta X_m(t), \delta Y_m(t), \delta q(t), \delta p(t)$, $\delta X_c(t), \delta Y_c(t)]^T$ is the vector operator of the system, $ n(t)=[\sqrt{2 \kappa_a} X_a^{\rm i n}(t), \sqrt{2 \kappa_a} Y_a^{\rm i n}(t), \sqrt{2 \kappa_m} X_m^{\rm i n}(t)$, $\sqrt{2 \kappa_m} Y_m^{\rm in}(t), 0, \xi(t), \sqrt{2 \kappa_c} X_c^{\rm in}(t), \sqrt{2 \kappa_c} Y_c^{\rm in}(t)]^T$ is the vector operator of the input noise, and
\begin{align}\label{eq6}
& A=\left(\begin{array}{cccccccc}
	-\kappa_a & \Delta_a & 0 & g_{\rm am} & 0 & 0 & 0 & 0 \\
	-\Delta_a & -\kappa_a & -g_{\rm am} & 0 & 0 & 0 & 0 & 0 \\
	0 & g_{\rm am} & -\kappa_m & U & G_m & 0 & 0 & 0 \\
	-g_{\rm am} & 0 & -V & -\kappa_m & 0 & 0 & 0 & 0 \\
	0 & 0 & 0 & 0 & 0 & \omega_b & 0 & 0 \\
	0 & 0 & 0 & -G_m & -\omega_b & -\gamma_b & 0 & -G_c \\
	0 & 0 & 0 & 0 & G_c & 0 & -\kappa_c & \tilde{\Delta}_c \\
	0 & 0 & 0 & 0 & 0 & 0 & -\tilde{\Delta}_c & -\kappa_c
\end{array}\right)
\end{align}
is the drift matrix, where $U=\tilde{\Delta}_m+3 \Delta_k$ and $V=(\tilde{\Delta}_m+\Delta_k)$. According to the Routh-Hurwitz criterion~\cite{dejesus1987routh}, when all eigenvalues of the matrix $A$ have negative real parts, the system is stable. In what follows, the stable condition is ensured numerically.

When the system is stable, it reaches a unique steady state, independently of the initial condition. Since quantum noises $\xi(t)$ and $\sigma_{\rm in}(t)$ are zero-mean quantum Gaussian noises and the dynamics is linearized [see Eq.~(\ref{eq4})], the quantum steady state for fluctuations is a zero-mean continuous variable Gaussian state, fully characterized by an $8\times8$ covariance matrix $\mathcal{V}_{lk}(\infty)=\frac{1}{2}\langle u_l(\infty) u_k\left(\infty\right)+u_k\left(\infty\right) u_l(\infty)\rangle(l,k=1,2, \ldots, 8)$.
The matrix $\mathcal{V}$ can be obtained by directly solving the Lyapunov
equation~\cite{vitali2007optomechanical} 
\begin{align}\label{eq7}
	A \mathcal{V}+\mathcal{V} A^T+\mathcal{D} =0,
\end{align}
 where $D=\operatorname{diag}\left[\kappa_a\left(2 n_a+1\right), \kappa_a\left(2 n_a+1\right), \kappa_m\left(2 n_m+1\right) \right.\\ \left., \kappa_m\left(2 n_m+1\right), 0, \gamma_b\left(2 n_b+1\right), \kappa_a\left(2 n_c+1\right), \kappa_a\left(2 n_c+1\right)\right]$ is defined by $\langle {n}_l(t) {n}_k(t^{\prime})$ $+{n}_k(t^{\prime}) {n}_l(t)\rangle=2\mathcal{D}_{lk} \delta(t-t^{\prime})$. Once the matrix $\mathcal{V}$ is obtained by solving Eq.~(\ref{eq7}), arbitrary bipartite entanglement can be quantified by the logarithmic negativity~\cite{vidal2002computable,adesso2004extremal,plenio2005logarithmic},
\begin{align}\label{eq10}
	E_N\equiv \max [0,-\ln 2{\eta^-}]
\end{align}
with
\begin{align}\label{eq21}
	\eta^-=2^{-1/2}[\Sigma-(\Sigma^2-4\rm det \mathcal{V}_4)^{1/2}]^{1/2},
\end{align}
where $ \Sigma=\rm{det} \mathcal{A}+\rm{det} \mathcal B-2\rm{det} \mathcal C $ and 
$\mathcal{V}_4=\begin{pmatrix}
	\mathcal A &\mathcal C\\
	\mathcal {C^T}& \mathcal B\\
\end{pmatrix}$
is the $4 \times4 $ block form of the correlation matrix, associated with two modes of interest~\cite{simon2000peres}. $\mathcal{A}$, $\mathcal{B}$, and $\mathcal{C}$ are the $ 2\times2 $ blocks of $\mathcal{V}_4$. A positive logarithmic negativity ($E_N > 0$) denotes the presence of bipartite entanglement of the two modes of interest in
the considered system. 
\section{nonreciprocal microwave-optical entanglement}\label{sec4}
\begin{figure}
	\includegraphics[scale=0.41]{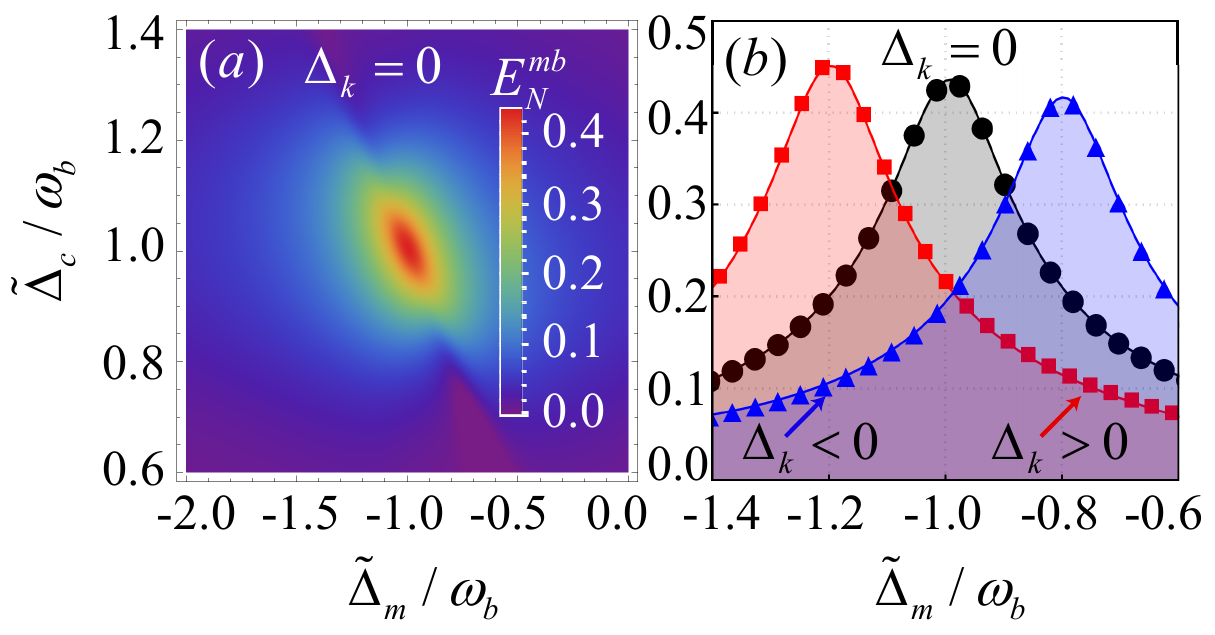}  
	\caption{(a) Density plot of the magnon-phonon entanglement $E_N^{\rm mb}$  vs the normalized parameters $\tilde{\Delta}_m/\omega_b$ and $\tilde{\Delta}_c/\omega_b$ without the magnon Kerr effect $\Delta_k=0$, where $\Delta_a=-\omega_b$. (b) The magnon-phonon entanglement $E_N^{\rm mb}$ vs the normalized parameter $\tilde{\Delta}_m/\omega_b$ with and without the magnon Kerr effect, where $\omega_a / 2 \pi=10$ GHz, $\omega_b / 2 \pi=40$ MHz,  $\omega_c/2\pi=190$ THz, $\Delta_a=-\omega_b$, $\tilde{\Delta}_c=\omega_b$, $\kappa_{a} / 2 \pi=\kappa_{m} / 2 \pi=1$ {MHz}, $\kappa_c / 2 \pi=2$ {MHz}, $\gamma_b / 2 \pi=100$ Hz, $g_{\rm am} / 2 \pi=4$ MHz, $G_m / 2 \pi=2$ MHz, $G_c / 2 \pi=8$ MHz, $T=10$ mK, $|\Delta_k|=0.1\omega_b$.}\label{fig2}
\end{figure}
In our setup, there are two kinds of microwave-optical entanglement, i.e., entanglement between photons in the optical cavity and photons in the microwave LC resonator ($E_N^{\rm ac}$) or magnons in the magnon mode ($E_N^{\rm mc}$). 

\begin{figure*}
	\includegraphics[scale=0.40]{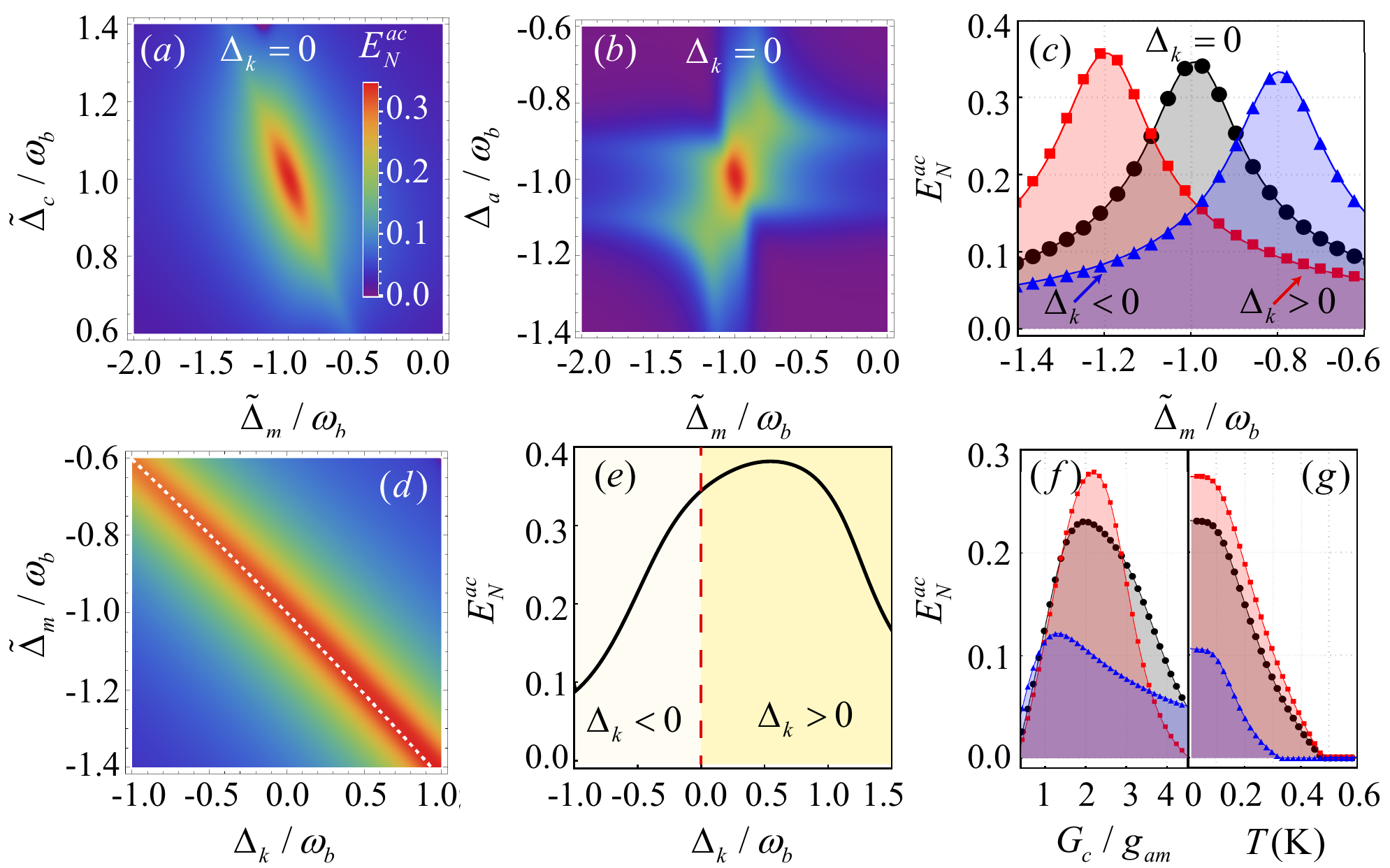}  
	\caption{Density plot of the microwave-optical photon-photon entanglement $E_N^{\rm ac}$  vs the normalized parameters (a) $\tilde{\Delta}_m/\omega_b$ and $\tilde{\Delta}_c/\omega_b$, (b) $\tilde{\Delta}_m/\omega_b$ and $\Delta_a/\omega_b$ without the magnon Kerr effect. (c) The microwave-optical photon-photon entanglement $E_N^{\rm ac}$ vs the normalized parameter $\tilde{\Delta}_m/\omega_b$ with and without the magnon Kerr effect. (d) Density plot of the microwave-optical photon-photon entanglement $E_N^{\rm ac}$ vs the normalized parameters $\tilde{\Delta}_m/\omega_b$ and $\Delta_k/\omega_b$. The microwave-optical photon-photon entanglement $E_N^{\rm ac}$ vs (e) $\Delta_k/\omega_b$, with $\tilde{\Delta}_m = -0.4 \Delta_k^2/\omega_b-2.06 \Delta_k-\omega_b$, (f) $G_c/g_{\rm am}$ with $\tilde{\Delta}_m/\omega_b=-1.11$, and (g) the bath temperature $T$ with $\tilde{\Delta}_m/\omega_b=-1.11$. In (a)-(g), $|\Delta_k|=0.1\omega_b$ and $\Delta_a=-\tilde{\Delta}_c=-\omega_b$ are taken. Other parameters are the same as those in Fig.~\ref{fig2}.
	}\label{fig3}
\end{figure*}

\subsection{Nonreciprocal magnon-phonon entanglement}
To study them, we operate the optical cavity at the red sideband ($\tilde{\Delta}_c=\omega_b$) and the magnon mode at the blue sideband ($\tilde{\Delta}_m=-\omega_b$). This is because, at the magnon blue sideband,  the Stokes scattering process is dominant, giving rise to entanglement between magnons and phonons. But to observe it, cooling mechanical resonator to the ground state is crucial, thus we need to operate the optical cavity at the red sideband to active the anti-Stokes scattering process, then the mechanical resonator can be significantly cooled to its ground state by transfer the thermal phonon numbers to the optical cavity. These two effects allow the optimal magnon-phonon entanglement observed, as demonstrated by Fig.~\ref{fig2}(a), in which the following feasible parameters are chosen~\cite{shen2022mechanical,potts2021dynamical} $\omega_a / 2 \pi=10$ GHz, $\omega_b / 2 \pi=40$ MHz,  $\omega_c/2\pi=190$ THz, $\kappa_{a} / 2 \pi=\kappa_{m} / 2 \pi=1$ {MHz}, $\kappa_c / 2 \pi=2$ {MHz}, $\gamma_b / 2 \pi=100$ Hz, $g_{\rm am} / 2 \pi=4$ MHz, $G_m / 2 \pi=2$ MHz, $G_c / 2 \pi=8$ MHz, $T=10$ mK.

In Fig.~\ref{fig2}(a), the magnon Kerr effect is not taken into account (i.e., $\Delta_k=0$). In fact, the magnon Kerr effect can give rise to nonreciprocal magnon-phonon entanglement, as illustrated by Fig.~\ref{fig2}(b), where the optical cavity is fixed at the red sideband ($\tilde{\Delta}_c=\omega_b$). Specifically, when the Kerr coefficient is possitive ($K_0>0$), the caused shift $\Delta_k>0$, leading to a left-shift for the magnon-phonon entanglement  (see the red line). But when $K_0<0$, we have $\Delta_k<0$, in this case we find the magnon-phonon entanglement has a right-shift (see the blue line).
In addition, the value of the maximal magnon-phonon entanglement also exhibits different behaviors when $\Delta_k$ is tuned from the positive ($\Delta_k>0$) to negative ($\Delta_k<0$). This differences come from the pair-magnon effect induced by linearization of the magnon Kerr effect. When $\Delta_k=0.1\omega_b>0$, the maximal value of the magnon-phonon entanglement is located at $\tilde{\Delta}_m\approx-1.2\omega_b$. But when $\Delta_k=-0.1\omega_b<0$, it is located at $\tilde{\Delta}_m\approx-0.8\omega_b$. These presice evaluation is given by $\tilde{\Delta}_m\approx-\omega_b-2\Delta_k$. This is because in the presence of the magnon Kerr effect, the effective frequency detuning indeed becomes  $\tilde{\Delta}_m+2\Delta_k$ [see the term related to $\delta m$ in Eq.~(\ref{eq4})].

{To explain why entanglement is enhanced (reduced) when $\Delta_k>0~(<0)$, one can see that the mean magnon number increases (decreases) when $\Delta_k>0~(<0)$, directly obtained from Eq.~(\ref{q5}). Specifically, given that $\Delta_a=-\omega_b$ and $\tilde{\Delta}_m<0$ are chosen [see Fig.~\ref{fig2}(b)], $\Delta_k \tilde{\Delta}_m>0$ (i.e., $\Delta_k<0$) results in a decrease in the mean magnon number. Conversely, when $\Delta_k \tilde{\Delta}_m<0$ (i.e., $\Delta_k>0$), the mean magnon number increases. A larger mean magnon number corresponds to a stronger effective magnomechanical coupling strength [see Eq.~(\ref{eq4})], giving rise to strong entanglement. Therefore, the sign of $\Delta_k$ directly influences the statistics of the mean magnon number, thereby indirectly enhancing or weakening the effective magnomechanical coupling strength for quantum entanglement generation.}

\begin{figure*}
	\includegraphics[scale=0.4]{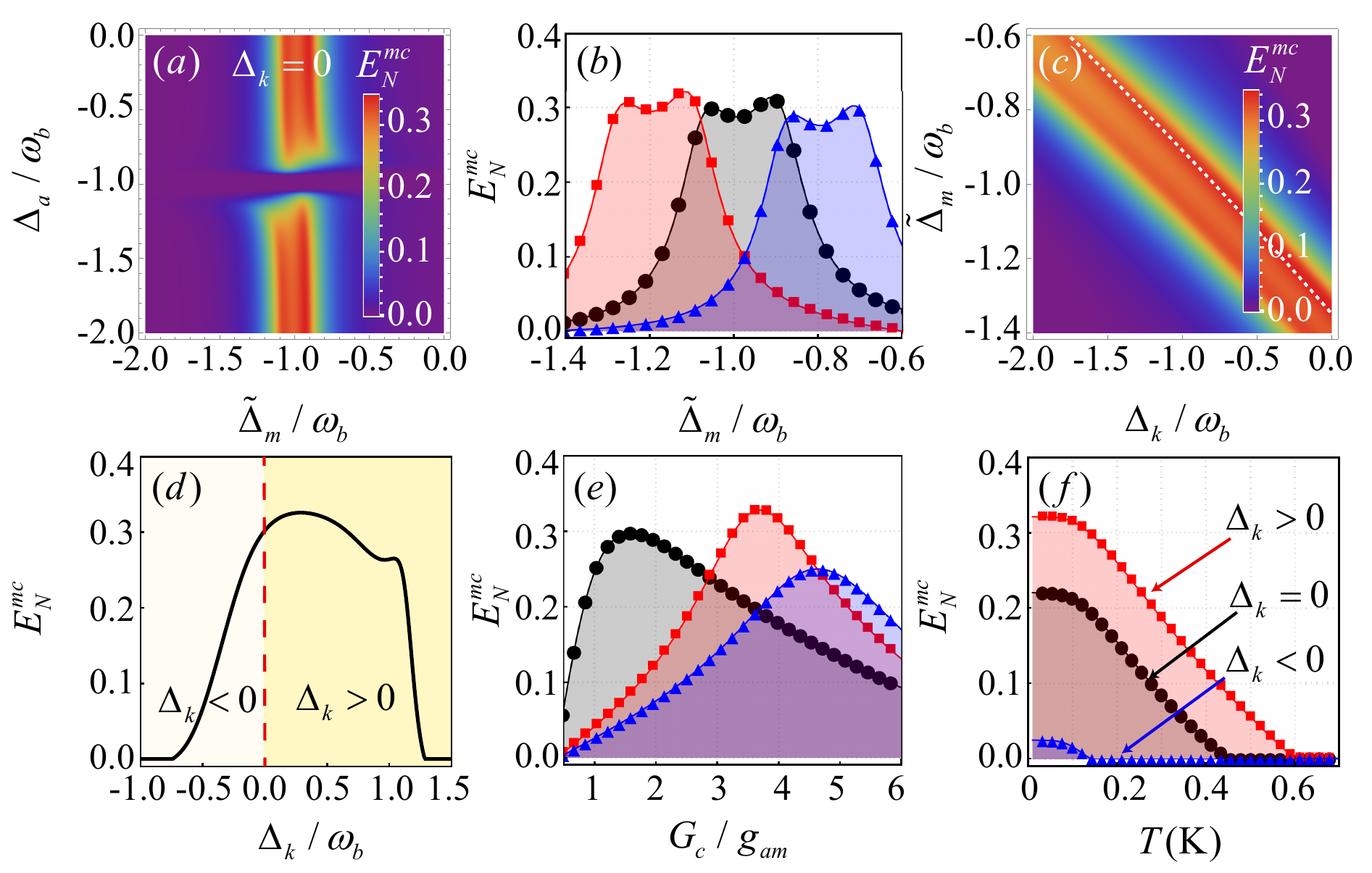}  
	\caption{Density plot of the optical photon-magnon entanglement $E_N^{\rm mc}$  vs the normalized parameters (a) $\tilde{\Delta}_m/\omega_b$ and $\tilde{\Delta}_c/\omega_b$, without the magnon Kerr effect. (b) The optical photon-magnon entanglement $E_N^{\rm mc}$ vs the normalized parameter $\tilde{\Delta}_m/\omega_b$ with and without the magnon Kerr effect. (c) Density plot of the optical photon-magnon entanglement $E_N^{\rm mc}$ vs the normalized parameters $\tilde{\Delta}_m/\omega_b$ and  $\Delta_k/\omega_b$. (d) The optical photon-magnon entanglement $E_N^{\rm mc}$ vs $\Delta_k/\omega_b$, with $\tilde{\Delta}_m = -0.45 \Delta_k^2/\omega_b-2.1 \Delta_k-0.91\omega_b$. The optical photon-magnon entanglement $E_N^{\rm mc}$ vs (e) the normalized parameter $G_c/g_{\rm am}$ with $\tilde{\Delta}_m/\omega_b=-1$, and (f) the bath temperature $T$ with $\tilde{\Delta}_m/\omega_b=-1.11$. In (a)-(f), $|\Delta_k|=0.1\omega_b$, $\Delta_a=0,\tilde{\Delta}_c=\omega_b$. Other parameters are the same as those in Fig.~\ref{fig2}.
	}\label{fig4}
\end{figure*}

\subsection{Nonreciprocal microwave-optical photon-photon entanglement}

Due to beam splitter interactions including the optical photon-phonon coupling and the microwave photon-magnon coupling, the established magnon-phonon entanglement in Fig.~\ref{fig2}(a) is distributed to both the subsystems of optomechanics and cavity magnonics, leading to indirect microwave-optical photon-photon entanglement. {When the {red-sideband} optical photons are resonant with the phonons}, while the microwave photons are resonant with the magnons, the microwave-optical photon-photon entanglement can be optimal, as clearly demonstrated by Figs.~\ref{fig3}(a) and \ref{fig3}(b). Figure ~\ref{fig3}(c) further illustrates the nonreciprocal behavior of the microwave-optical photon-photon entanglement vs the normalized magnon detuning $\tilde{\Delta}_m/\omega_b$ induced by the magnon Kerr effect, where $\Delta_a=-\omega_b$ and $\tilde{\Delta}_c=\omega_b$ are fixed. One can obviously see that the microwave-optical photon-photon entanglement has a similar behavior with the magnon-phonon entanglement in Fig.~\ref{fig2}(b) when the magnon Kerr coefficient is tuned from negative to positive. We also examine the behavior of the microwave-optical photon-photon entanglement with the normalized parameters $\tilde{\Delta}_m/\omega_b$ and $\Delta_k/\omega_b$ in Fig.~\ref{fig3}(d). We find that the optimal microwave-optical photon-photon entanglement is realized only when $\tilde{\Delta}_m \simeq -0.4 \Delta_k^2/\omega_b-2.06 \Delta_k-\omega_b$ is satisfied (see the white dashed line). When this condition is satisfied, we show that the optimal microwave-optical photon-photon entanglement can be enhanced in the region of $\Delta_k>0$ compared to the case without the magnon Kerr effect ($\Delta_k=0$). However, in the region of $\Delta_k<0$, the optimal microwave-optical photon-photon entanglement is significantly weakened. Moreover, the optimal microwave-optical photon-photon entanglement decreases monotonously with $\Delta_k$ in the region of $\Delta_k<0$, while it increases first and then decreases with $\Delta_k$ in the region of $\Delta_k>0$, as illustrated in Fig.~\ref{fig3}(e). In fact, the microwave-optical photon-photon entanglement can also be tuned by varying the linearized optomechanical coupling $G_c$, as demonstrated by Fig.~\ref{fig3}(f). By introducing the magnon Kerr effect, we show that the optimal microwave-optical photon-photon entanglement can be nonreciprocally enhanced. At the same time, we find that the microwave-optical photon-photon entanglement can be nonreciprocally robust against the bath temperature [see Fig.~\ref{fig3}(g)]. More importantly, the introduced magnon Kerr effect can significantly improve the survival temperature of the microwave-optical photon-photon entanglement.

\subsection{Nonreciprocal optical photon-magnon entanglement}

When the magnon-phonon entanglement is generated, the phonon mediated optical photon-magnon entanglement can also be produced because the beam splitter interaction between the optical photons and phonons allows the entanglement redistributed to the subsystem of the optomechanics. To produce nonreciprocal optical photon-magnon entanglement, the magnon Kerr effect is introduced. Compared Fig.~\ref{fig4}(a) with Fig.~\ref{fig3}(b), the complementary entanglement distribution clearly shows that breaking the microwave photon-magnon resonance ($\Delta_a \neq \tilde{\Delta}_m$) is benifical to generate the optical photon-magnon entanglement. Figure~\ref{fig4}(b) shows that the optical photon-magnon entanglement has opposite behaviors when the sign of the Kerr coefficient is opposite. To obtain the optimal entanglement in the presence of the magnon Kerr effect, the condition $\tilde{\Delta}_m \simeq -0.45 \Delta_k^2/\omega_b-2.1 \Delta_k-0.91\omega_b$ must be approximately satisfied, which is given by the white dashed curve in Fig. ~\ref{fig4}(c). When this optimal condition is kept, the optimal optical photon-magnon entanglement has the similar behavior with the optimal microwave-optical photon-photon entanglement in Fig.~\ref{fig3}(e) by tuning the Kerr coefficient from negative (opsitive) to oppsitive (negative), as demonstrated by Fig.~\ref{fig4}(d). Obviously, the optimal optical photon-magnon entanglement decreases with increasing $\Delta_k$ in the region of $\Delta_k<0$. However, in the region of $\Delta_k>0$, the optimal optical photon-magnon entanglement increases first and then decreases with increasing $\Delta_k$.  This indicates that the optical photon-magnon entanglement can be nonreciprocally enhanced by the magnon Kerr effect. Figure~\ref{fig4}(e) further shows that the optical photon-magnon entanglement can be tuned by the linearized optomechanical coupling strength. Besides, by introducing the magnon Kerr effect, not only the optical photon-magnon entanglement has nonreciprocal behaviors, but also a stronger optomechanical coupling is required to obtain the optimal optical photon-magnon entanglement. Figure~\ref{fig4}(f) indicates that nonreciprocity induced by the magnon Kerr effect not only can lead to the optical photon-magnon entanglement more robust against the bath temperature, but also it can improve the entanglement survival temperature. 
\begin{figure}
	\includegraphics[scale=0.4]{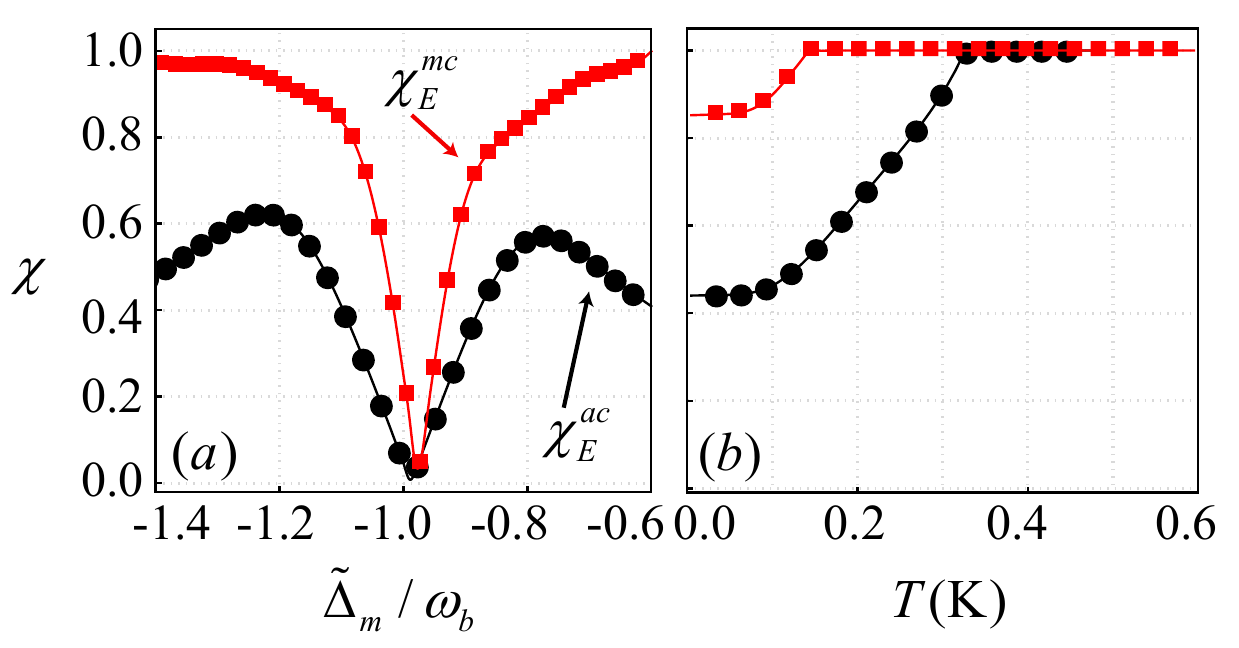}  
	\caption{Bidirectional contrast ratio $\chi$ vs the normalized parameters (a) $\tilde{\Delta}_m/\omega_b$ and (b) the bath temperature $T$, where $|\Delta_k|=0.1\omega_b$. Other parameters are the same as those in Fig.~\ref{fig2}.	}\label{fig5}
\end{figure}

\subsection{Bidirectional contrast ratio}

To quantitatively describe nonreciprocal entanglement induced by the magnon Kerr effect, we introduce the bidirectional contrast ratio $\chi$ (satisfying 0 $\leq \chi \leq$ 1) for bipartite entanglement to show nonreciprocity, i.e.,
\begin{align}
	\chi_E^{i j} =\frac{\left|E_N^{i j}(>0)-E_N^{i j}(<0)\right|}{E_N^{i j}(>0)+E_N^{i j}(<0)}. 
\end{align}
The parameter $\chi_E^{i j}$ equals $1$ for ideal nonreciprocity and $0$ for no nonreciprocity in bipartite entanglement. A higher contrast ratio $\chi$ indicates stronger nonreciprocity. To clarify this, we numerically plot the bidirectional contrast ratio $\chi$ as a function of frequency detuning $\tilde{\Delta}_m$ in Fig.~\ref{fig5}(a). The figure demonstrates that by tuning $\tilde{\Delta}_m$, the nonreciprocity of the microwave-optical photon-photon entanglement and the optical photon-mangon entanglement can be switched on or off. Moreover, the bidirectional contrast ratio for both types of entanglement can be adjusted from 0 to 1 by varying $\tilde{\Delta}_m$, indicating that ideal nonreciprocity for the microwave-optical photon-photon entanglement and the optical photon-mangon entanglement can be achieved by controlling the magnon frequency detuning.

The effect of the bath temperature on the bidirectional contrast is further examined in Fig.~\ref{fig5}(b). The results show that the nonreciprocity of the microwave-optical photon-photon entanglement and the optical photon-mangon entanglement gradually increases with increasing the bath temperature. Interestingly, ideal nonreciprocity for both entanglements can be realized at higher temperatures, indicating that elevated temperatures are beneficial for achieving optimal nonreciprocal entanglement.

\section{conclusion}\label{s5}

In summary, we propose a novel scheme that leverages the magnon Kerr effect to generate and control nonreciprocal microwave-optical entanglement in cavity optomagnomechanical (COMM) systems. By utilizing the strong coupling between a magnonic resonator, microwave cavity, and optical cavity, we demonstrate how the direction of the external magnetic field can be exploited to induce nonreciprocal behavior in entanglement dynamics. We show that tuning parameters, such as magnon frequency detuning, allows for enhanced nonreciprocal entanglement across multiple domains, including magnon-phonon, microwave-optical, and photon-magnon entanglements. Moreover, our findings highlight the potential for achieving ideal nonreciprocity in microwave-optical and photon-magnon entanglements, which is crucial for quantum information processing and quantum networks. This work provides a promising path for designing nonreciprocal quantum devices capable of operating across both microwave and optical bands, contributing to the development of hybrid quantum systems and advanced quantum technologies.

%\section{ ACKNOWLEDGMENTS}
WX is supported by the Natural Science Foundation of Zhejiang Province (Grant No. LY24A040004), the "Pioneer" and "Leading Goose" R\&D Program of Zhejiang (Grant No. 2025C01028), and the Shenzhen International Quantum Academy (Grant No. SIQA2024KFKT010). YWW is supported by the Natural Science Foundation of Zhejiang Province (Grant No.~LY23A40002) and Wenzhou Science and Technology Plan Project~(No.~L20240004). 

\bibliographystyle{iopart-num} 
\bibliography{ms}

\providecommand{\newblock}{}
\begin{thebibliography}{10}
\expandafter\ifx\csname url\endcsname\relax
  \def\url#1{{\tt #1}}\fi
\expandafter\ifx\csname urlprefix\endcsname\relax\def\urlprefix{URL }\fi
\providecommand{\eprint}[2][]{\url{#2}}

\bibitem{rameshti2022cavity}
Rameshti B~Z, Kusminskiy S~V, Haigh J~A, Usami K, Lachance-Quirion D, Nakamura
  Y, Hu C~M, Tang H~X, Bauer G~E and Blanter Y~M 2022 {\em Phys. Rep.\/} {\bf
  979} 1--61

\bibitem{yuan2022quantum}
Yuan H, Cao Y, Kamra A, Duine R~A and Yan P 2022 {\em Phys. Rep.\/} {\bf 965}
  1--74

\bibitem{prabhakar2009spin}
Prabhakar A and Stancil D~D 2009 {\em Spin waves: Theory and applications\/}
  (Springer)

\bibitem{van1958spin}
Van~Kranendonk J and Van~Vleck J 1958 {\em Rev. Mod. Phys.\/} {\bf 30} 1

\bibitem{schmidt2020ultra}
Schmidt G, Hauser C, Trempler P, Paleschke M and Papaioannou E~T 2020 {\em
  phys. status solidi (b)\/} {\bf 257} 1900644

\bibitem{mallmann2013yttrium}
Mallmann E, Sombra A, Goes J and Fechine P 2013 {\em Solid State Phenom.\/}
  {\bf 202} 65--96

\bibitem{geller1957crystal}
Geller S and Gilleo M 1957 {\em J. Phys. Chem. Solids\/} {\bf 3} 30--36

\bibitem{li2020hybrid}
Li Y, Zhang W, Tyberkevych V, Kwok W~K, Hoffmann A and Novosad V 2020 {\em J.
  Appl. Phys.\/} {\bf 128} 130902

\bibitem{huebl2013high}
Huebl H, Zollitsch C~W, Lotze J, Hocke F, Greifenstein M, Marx A, Gross R and
  Goennenwein S~T 2013 {\em Phys. Rev. Lett.\/} {\bf 111} 127003

\bibitem{tabuchi2014hybridizing}
Tabuchi Y, Ishino S, Ishikawa T, Yamazaki R, Usami K and Nakamura Y 2014 {\em
  Phys. Rev. Lett.\/} {\bf 113} 083603

\bibitem{zhang2014strongly}
Zhang X, Zou C~L, Jiang L and Tang H~X 2014 {\em Phys. Rev. Lett.\/} {\bf 113}
  156401

\bibitem{zhang2015magnon}
Zhang X, Zou C~L, Zhu N, Marquardt F, Jiang L and Tang H~X 2015 {\em Nat.
  Commun.\/} {\bf 6} 8914

\bibitem{zhang2017observation}
Zhang D, Luo X~Q, Wang Y~P, Li T~F and You J 2017 {\em Nat. Commun.\/} {\bf 8}
  1368

\bibitem{zhang2019experimental}
Zhang X, Ding K, Zhou X, Xu J and Jin D 2019 {\em Phys. Rev. Lett.\/} {\bf 123}
  237202

\bibitem{zhao2020observation}
Zhao J, Liu Y, Wu L, Duan C~K, Liu Y~X and Du J 2020 {\em Phys. Rev. Appl.\/}
  {\bf 13}(1) 014053

\bibitem{sadovnikov2022exceptional}
Sadovnikov A~V, Zyablovsky A~A, Dorofeenko A~V and Nikitov S~A 2022 {\em Phys.
  Rev. Appl.\/} {\bf 18} 024073

\bibitem{liu2019observation}
Liu H, Sun D, Zhang C, Groesbeck M, Mclaughlin R and Vardeny Z~V 2019 {\em Sci.
  Adv.\/} {\bf 5} eaax9144

\bibitem{cao2019exceptional}
Cao Y and Yan P 2019 {\em Phys. Rev. B\/} {\bf 99} 214415

\bibitem{yang2021controlling}
Yang Z~B, Liu X~D, Yin X~Y, Ming Y, Liu H~Y and Yang R~C 2021 {\em Phys. Rev.
  Appl.\/} {\bf 15} 024042

\bibitem{wang2019nonreciprocity}
Wang Y~P, Rao J, Yang Y, Xu P~C, Gui Y, Yao B, You J and Hu C~M 2019 {\em Phys.
  Rev. Lett.\/} {\bf 123} 127202

\bibitem{harder2021coherent}
Harder M, Yao B, Gui Y and Hu C~M 2021 {\em J. Appl. Phys.\/} {\bf 129} 201101

\bibitem{hei2021enhancing}
Hei X~L, Dong X~L, Chen J~Q, Shen C~P, Qiao Y~F and Li P~B 2021 {\em Phys. Rev.
  A\/} {\bf 103} 043706

\bibitem{rao2021interferometric}
Rao J, Xu P, Gui Y, Wang Y, Yang Y, Yao B, Dietrich J, Bridges G, Fan X, Xue D
  {\em et~al.\/} 2021 {\em Nat. Commun.\/} {\bf 12} 1933

\bibitem{xu2023quantum}
Xu D, Gu X~K, Li H~K, Weng Y~C, Wang Y~P, Li J, Wang H, Zhu S~Y and You J 2023
  {\em Phys. Rev. Lett.\/} {\bf 130} 193603

\bibitem{xu2024macroscopic}
Xu D, Gu X~K, Weng Y~C, Li H~K, Wang Y~P, Zhu S~Y and You J 2024 {\em Quantum
  Science and Technology\/} {\bf 9} 035002

\bibitem{zuo2024cavity}
Zuo X, Fan Z~Y, Qian H, Ding M~S, Tan H, Xiong H and Li J 2024 {\em New J.
  Phys.\/} {\bf 26} 031201

\bibitem{aspelmeyer2014cavity}
Aspelmeyer M, Kippenberg T~J and Marquardt F 2014 {\em Rev. Mod. Phys.\/} {\bf
  86} 1391--1452

\bibitem{vitali2007optomechanical}
Vitali D, Gigan S, Ferreira A, B{\"o}hm H, Tombesi P, Guerreiro A, Vedral V,
  Zeilinger ~f~A and Aspelmeyer M 2007 {\em Phys. Rev. Lett.\/} {\bf 98} 030405

\bibitem{xiong2022higher}
Xiong W, Li Z, Zhang G~Q, Wang M, Li H~C, Luo X~Q and Chen J 2022 {\em Phys.
  Rev. A\/} {\bf 106} 033518

\bibitem{xiong2021higher}
Xiong W, Li Z, Song Y, Chen J, Zhang G~Q and Wang M 2021 {\em Phys. Rev. A\/}
  {\bf 104} 063508

\bibitem{xiong2016cross}
Xiong W, Jin D~Y, Qiu Y, Lam C~H and You J 2016 {\em Phys. Rev. A\/} {\bf 93}
  023844

\bibitem{kani2022intensive}
Kani A, Sarma B and Twamley J 2022 {\em Phys. Rev. Lett.\/} {\bf 128} 013602

\bibitem{li2018magnon}
Li J, Zhu S~Y and Agarwal G 2018 {\em Phys. Rev. Lett.\/} {\bf 121} 203601

\bibitem{li2019squeezed}
Li J, Zhu S~Y and Agarwal G 2019 {\em Phys. Rev. A\/} {\bf 99} 021801

\bibitem{lu2021exceptional}
Lu T~X, Zhang H, Zhang Q and Jing H 2021 {\em Phys. Rev. A\/} {\bf 103} 063708

\bibitem{he2024mechanical}
He W~D, Fan X~H, Liu M~Y, Zhang G~Q, Li H~C and Xiong W 2024 {\em Adv. Quantum
  Tech.\/}  2400275

\bibitem{fan2023microwave}
Fan Z~Y, Qiu L, Gr{\"o}blacher S and Li J 2023 {\em Laser \& Photonics
  Reviews\/} {\bf 17} 2200866

\bibitem{fan2022optical}
Fan Z~Y, Shen R~C, Wang Y~P, Li J and You J 2022 {\em Phys. Rev. A\/} {\bf 105}
  033507

\bibitem{zhang2019theory}
Zhang G, Wang Y and You J 2019 {\em Science China Physics, Mechanics \&
  Astronomy\/} {\bf 62} 1--11

\bibitem{wang2016magnon}
Wang Y~P, Zhang G~Q, Zhang D, Luo X~Q, Xiong W, Wang S~P, Li T~F, Hu C~M and
  You J 2016 {\em Phys. Rev. B\/} {\bf 94} 224410

\bibitem{wang2018bistability}
Wang Y~P, Zhang G~Q, Zhang D, Li T~F, Hu C~M and You J 2018 {\em Phys. Rev.
  Lett.\/} {\bf 120} 057202

\bibitem{zheng2023tutorial}
Zheng S, Wang Z, Wang Y, Sun F, He Q, Yan P and Yuan H 2023 {\em J. Appl.
  Phys.\/} {\bf 134}

\bibitem{shen2022mechanical}
Shen R~C, Li J, Fan Z~Y, Wang Y~P and You J 2022 {\em Phys. Rev. Lett.\/} {\bf
  129} 123601

\bibitem{bi2021tristability}
Bi M, Yan X, Zhang Y and Xiao Y 2021 {\em Phys. Rev. B\/} {\bf 103} 104411

\bibitem{xiong2022strong}
Xiong W, Tian M, Zhang G~Q and You J 2022 {\em Phys. Rev. B\/} {\bf 105} 245310

\bibitem{xiong2023optomechanical}
Xiong W, Wang M, Zhang G~Q and Chen J 2023 {\em Phys. Rev. A\/} {\bf 107}
  033516

\bibitem{peng2023strong}
Peng M~L, Tian M, Chen X~C, Zhang G~Q, Li H~C and Xiong W 2023 {\em arXiv
  e-prints\/}  arXiv--2304

\bibitem{zhang2021parity}
Zhang G~Q, Chen Z, Xiong W, Lam C~H and You J 2021 {\em Phys. Rev. B\/} {\bf
  104} 064423

\bibitem{liu2023switchable}
Liu G, Xiong W and Ying Z~J 2023 {\em Phys. Rev. A\/} {\bf 108}(3) 033704

\bibitem{chen2024nonreciprocal}
Chen J, Fan X~G, Xiong W, Wang D and Ye L 2024 {\em Phys. Rev. A\/} {\bf 109}
  043512

\bibitem{chen2023nonreciprocal}
Chen J, Fan X~G, Xiong W, Wang D and Ye L 2023 {\em Phys. Rev. B\/} {\bf 108}
  024105

\bibitem{zhong2020entanglement}
Zhong C, Han X, Tang H~X and Jiang L 2020 {\em Physical Review A\/} {\bf 101}
  032345

\bibitem{barzanjeh2012reversible}
Barzanjeh S, Abdi M, Milburn G~J, Tombesi P and Vitali D 2012 {\em Physical
  Review Letters\/} {\bf 109} 130503

\bibitem{wu2021deterministic}
Wu J, Cui C, Fan L and Zhuang Q 2021 {\em Physical Review Applied\/} {\bf 16}
  064044

\bibitem{zhong2020proposal}
Zhong C, Wang Z, Zou C, Zhang M, Han X, Fu W, Xu M, Shankar S, Devoret M~H,
  Tang H~X {\em et~al.\/} 2020 {\em Physical review letters\/} {\bf 124} 010511

\bibitem{meesala2024quantum}
Meesala S, Lake D, Wood S, Chiappina P, Zhong C, Beyer A~D, Shaw M~D, Jiang L
  and Painter O 2024 {\em Physical Review X\/} {\bf 14} 031055

\bibitem{wei2022tunable}
Wei T, Wu D, Miao Q, Yang C and Luo J 2022 {\em Optics Express\/} {\bf 30}
  10135--10151

\bibitem{sahu2023entangling}
Sahu R, Qiu L, Hease W, Arnold G, Minoguchi Y, Rabl P and Fink J~M 2023 {\em
  Science\/} {\bf 380} 718

\bibitem{zheng2024nonreciprocal}
Zheng Q, Zhong W, Cheng G and Chen A 2024 {\em Journal of Applied Physics\/}
  {\bf 135}

\bibitem{zuo2023cavity}
Zuo X, Fan Z~Y, Qian H, Ding M~S, Tan H, Xiong H and Li J 2023 {\em arXiv
  preprint arXiv: 2310.19237\/}

\bibitem{li2025microwave}
Li H~T, Fan Z~Y, Zhu H~B, Gr{\"o}blacher S and Li J 2025 {\em Laser \&
  Photonics Reviews\/}  2401348

\bibitem{fan2024nonreciprocal}
Fan X~H, Zhang Y~N, Yu J~P, Liu M~Y, He W~D, Li H~C and Xiong W 2024 {\em Adv.
  Quantum Tech.\/} {\bf 7} 2400043

\bibitem{jiao2020nonreciprocal}
Jiao Y~F, Zhang S~D, Zhang Y~L, Miranowicz A, Kuang L~M and Jing H 2020 {\em
  Phys. Rev. Lett.\/} {\bf 125} 143605

\bibitem{maayani2018flying}
Maayani S, Dahan R, Kligerman Y, Moses E, Hassan A~U, Jing H, Nori F,
  Christodoulides D~N and Carmon T 2018 {\em Nature\/} {\bf 558} 569--572

\bibitem{malykin2000sagnac}
Malykin G~B 2000 {\em Physics-Uspekhi\/} {\bf 43} 1229

\bibitem{schoelkopf2008wiring}
Schoelkopf R and Girvin S 2008 {\em Nature\/} {\bf 451} 664--669

\bibitem{pirandola2016physics}
Pirandola S and Braunstein S~L 2016 {\em Nature\/} {\bf 532} 169--171

\bibitem{krastanov2021optically}
Krastanov S, Raniwala H, Holzgrafe J, Jacobs K, Lon{\v{c}}ar M, Reagor M~J and
  Englund D~R 2021 {\em Phys. Rev. Lett.\/} {\bf 127} 040503

\bibitem{agusti2022long}
Agust{\'\i} J, Minoguchi Y, Fink J~M and Rabl P 2022 {\em Phys. Rev. A\/} {\bf
  105} 062454

\bibitem{xiang2013hybrid}
Xiang Z~L, Ashhab S, You J and Nori F 2013 {\em Rev. Mod. Phys.\/} {\bf 85}
  623--653

\bibitem{clerk2020hybrid}
Clerk A, Lehnert K, Bertet P, Petta J and Nakamura Y 2020 {\em Nat. Phys.\/}
  {\bf 16} 257--267

\bibitem{zhang2016cavity}
Zhang X, Zou C~L, Jiang L and Tang H~X 2016 {\em Sci. Adv.\/} {\bf 2} e1501286

\bibitem{benguria1981quantum}
Benguria R and Kac M 1981 {\em Phys. Rev. Lett.\/} {\bf 46} 1

\bibitem{giovannetti2001phase}
Giovannetti V and Vitali D 2001 {\em Phys. Rev. A\/} {\bf 63} 023812

\bibitem{dejesus1987routh}
DeJesus E~X and Kaufman C 1987 {\em Phys. Rev. A\/} {\bf 35} 5288

\bibitem{vidal2002computable}
Vidal G and Werner R~F 2002 {\em Phys. Rev. A\/} {\bf 65} 032314

\bibitem{adesso2004extremal}
Adesso G, Serafini A and Illuminati F 2004 {\em Phys. Rev. A\/} {\bf 70} 022318

\bibitem{plenio2005logarithmic}
Plenio M~B 2005 {\em Phys. Rev. Lett.\/} {\bf 95} 090503

\bibitem{simon2000peres}
Simon R 2000 {\em Phys. Rev. Lett.\/} {\bf 84} 2726

\bibitem{potts2021dynamical}
Potts C~A, Varga E, Bittencourt V~A, Kusminskiy S~V and Davis J~P 2021 {\em
  Phys. Rev. X\/} {\bf 11} 031053

\end{thebibliography}

\end{document}